\documentclass{article}

\title{Trihamiltonian extensions of separable systems in the plane}
\author{Luca Degiovanni
\footnote{University of Torino, Mathematics Department}
}
\date{}

\usepackage{latexsym}
\usepackage{amssymb}
\newcommand{\Doppio}[1]{\mathbb{#1}}

\newcommand{\Reali}{\Doppio{R}}

\newcommand{\BD}{\begin{displaymath}} 
\newcommand{\ED}{\end{displaymath}}
\newcommand{\BE}{\begin{equation}}
\newcommand{\EE}{\end{equation}}
\newcommand{\BAS}{\begin{eqnarray*}}
\newcommand{\EAS}{\end{eqnarray*}}
\newcommand{\BA}{\begin{eqnarray}}
\newcommand{\EA}{\end{eqnarray}}

\newcommand{\D}{\mathrm{d}}

\newcommand{\Mu}{M}

\newcommand{\Pp}{P}
\newcommand{\Q}{Q}
\newcommand{\R}{R}
\newcommand{\Qd}{Q_d}
\newcommand{\Rd}{R_d}
\newcommand{\Xq}{X_Q}
\newcommand{\Xr}{X_R}

\newcommand{\half}{\frac{1}{2}}
\newcommand{\identity}{\mathrm{1\kern-3pt{I}}}

\newcommand{\DLie}[2]{\mathrm{L}_{#1}\,#2}

\newtheorem{Prop}{Proposition}

\newtheorem{Fatto}[Prop]{Fact}

\newcommand{\FrecciaAnd}[1]{
\put(0,0){\vector(3,2){15}}
\put(2,6){\tiny #1}
}

\newcommand{\FrecciaRit}[1]{
\put(15,0){\vector(-3,2){15}}
\put(8,6){\tiny #1}
}

\newcommand{\FrecciaGiu}[1]{
\put(7,19){\vector(0,-1){18}}
\put(8,8){\tiny #1}
}

\newcommand{\FrecciaP}{
\FrecciaAnd{$P$}
}

\newcommand{\FrecciaQ}{
\FrecciaRit{$Q$}
}

\newcommand{\FrecciaR}{
\FrecciaGiu{$R$}
}

\newcommand{\Scatola}[1]{
\put(0,0){\makebox(15,15){#1}}
}

\newcommand{\Nucleo}[1]{
\put(0,0){\FrecciaP}
\put(30,0){\FrecciaQ}
\put(15,25){\FrecciaR}
\put(15,10){\Scatola{#1}}
}

\begin{document}

\maketitle
\section*{Abstract}
A method to construct trihamiltonian extensions of a separable system is presented. The procedure is tested for systems, with a natural Hamiltonian, separable in classical sense in one of the four orthogonal separable coordinate systems of the Euclidean plane, and some explicit examples are constructed. Finally a conjecture on possible generalizations to other classes of systems is discussed: in particular, the method can be easily adapted to the eleven orthogonal separable coordinate sets of the Euclidean three-space.

\section{Introduction}
Separation of variables for Hamilton--Jacobi equation is a very effective method to find solutions of Hamiltonian systems. A set of coordinates in which a certain Hamilton--Jacobi equation can be separated is called a separable set of coordinates for that Hamiltonian system. The classical characterization of separability is restricted to the particular class of Hamiltonians defined on the cotangent bundle of a Riemannian manifold $(Q,g)$ with the form:
$$
H=\half g^{ij}p_i p_j + V(q)=G(q,p)+V(q)\,.
$$
This kind of Hamiltonians is usually called \emph{natural}. Moreover, in the classical theory, the separation is always performed through a \emph{contact transformation of coordinates}, i.e. a fibred symplectic transformation on $T^\ast Q$ obtained from a transformation in the configuration space $Q$.  In this contest a natural Hamiltonian is separable in a given set of coordinates only if its geodesic part $G$ is separable \cite{BenGen}. It is therefore possible to determine the sets of coordinates in which a certain Hamiltonian may be separated through the study of all separable coordinate systems associated to the metric $g$. The particular case in which the separation is performed in an orthogonal set of coordinates is called \emph{orthogonal separation}, other cases are said of \emph{general separation}. In recent years a rich and coordinates-indipendent characterization of both orthogonal and general separation was developed (see \cite{KeM1,KeM2,BenGen} and references therein).

The case of orthogonal separation is quite important not only because it is simpler, but also because it has been proved that in a Riemannian manifold of constant curvature only orthogonal separation is possible (see \cite{BenGen} for a review of such classical results). The classical analysis of orthogonally separable systems is due to St\"ackel and Eisenhart \cite{SepClass}. The central key of their analysis is the concept of \emph{St\"ackel matrix}, i.e. a $n\times n$ invertible matrix $S$ (where $n$ is the dimension of the Riemannian manifold) such that each element $S_i^j$ belonging to the $i$-th row of the matrix depends only on the coordinate $q_i$. St\"ackel proved that a geodetic Hamiltonian is orthogonally separable if and only if the nonvanishing controvariant components of the metric tensor, $g^{ii}$, form a row of the inverse of a St\"ackel matrix:
$$
g^{ii}=(S^{-1})^i_j \mbox{ for a fixed } j\,.
$$
The existence of a St\"ackel matrix is equivalent to the existence of a family of $n-1$ particular Killing tensors $K_{(l)}$ for the metric $g$. The vector space spanned by this family and the metric is the \emph{Killing--St\"ackel algebra} associated to the orthogonal coordinate systems \cite{Cha2001}. Vice versa, from a Killing--St\"ackel algebra it is possible to reconstruct the St\"ackel matrix of the coordinate system. Finally, also the separability of the complete natural Hamiltonians can be related to the St\"ackel matrix: the Killing--St\"ackel algebra allows to construct a family of $n-1$ potentials $V_{(l)}$ such that the Hamiltonians
$$
H_{(l)}= K_{(l)}^{ij}p_ip_j +V_{(l)}
$$
are in involution both mutually and with $H$. An alternative way to restate these results is the following: a family of $n$ Hamiltonians $H_i$ is orthogonally separable if and only if there exists a  St\"ackel matrix $S$ and a set of $n$ functions $\Phi_i(q_i,p_i)$, each depending only on the $i$-th coordinate, such that
\begin{equation}\label{SepStackel}
S\left(
\begin{array}{c}
H_1 \\
\vdots \\
H_n
\end{array}
\right)=\left(
\begin{array}{c}
\Phi_1(q_1,p_1) \\
\vdots \\
\Phi_n(q_n,p_n)
\end{array}
\right)
\end{equation}

These classical results has been put in a more handy form in \cite{Ben1997}, where it has been proved that a natural Hamiltonian is orthogonally separable if and only if exists a Killing tensor $K$ with $n$ normal and simple eigenvalues such that $d(KdV)=0$. A stronger result can be proved if $Q$ is also equipped with a \emph{conformal Killing tensor} with vanishing \emph{Nijenhuis torsion}, i.e. a tensor $L$ of type $(1,1)$ satisfying
$$
[LX,LY]-L[LX,Y]-L[X,LY]+L^2[X,Y]=0
$$
for all vector fields $X$ and $Y$ on $Q$, and
$$
\{L^{ij}p_ip_j,g^{ij}p_ip_j\}=c\,g^{ij}p_ip_j
$$
In this case it is indeed possible to generate the whole Killing--St\"ackel algebra by a recursion formula. These systems are called \emph{Benenti systems} or \emph{$L$-systems}. The recurrence property of Benenti systems was explained in the paper \cite{Mag2000}, where an extension of the systems with a Killing--St\"ackel algebra generated by the conformal Killing tensor $L$ is constructed, and it is proved that this extension fits in a bihamiltonian hierarchy defined on an extended manifold that contains the original phase space $T^\ast Q$ as a symplectic leaf of a degenerate Poisson structure. Therefore the recursion formula for the Killing tensor in the Killing--St\"ackel algebra is obtained from the bihamiltonian recursion scheme.

So far the only contact transformations has been considered in order to perform a separation of variables. In \cite{Skl1995} Sklyanin proposed to generalize this framework allowing general symplectic transformations and substituting the separation relations expressed by (\ref{SepStackel}) with the $n$ general equations:
$$
\left\{
\begin{array}{rcl}
W_1(q_1,p_1;\{H_i\}) &=& 0 \\
&\vdots\\
W_n(q_n,p_n;\{H_i\}) &=& 0
\end{array}
\right.
$$
Also this more general kind of separation has been interpreted in the bihamiltonian framework \cite{BiHamSep}. A different point of view could be found in the review \cite{Bla2000}.

The aim of the present work is to introduce, in the simple case of the Euclidean plane, a recipe for constructing trihamiltonian (and consequently bihamiltonian) extensions of a orthogonally separable natural Hamiltonian. This procedure works also for systems separable in symmetric sets of coordinates, whose Killing--St\"ackel algebra is in general not constructible from a conformal Killing tensor $L$.  Although the method seems promising also for the separable coordinate systems of the Euclidean three-space, the possibility of its extension to any orthogonally separable system remains unfortunately to be proved in general.

The construction presented in this paper in not interesting just as a new class example of trihamiltonian systems, in which the third Poisson structure links together different Lenard chains. Indeed it gives also an Hamiltonian interpretation of the concept of separation curves presented in \cite{Bla2000}: this curve is obtained through a reduction procedure from the common Casimir function associated to the trihamiltonian structure. Moreover starting from a trihamiltonian extension of a separable system in the plane it is possible to construct two different bihamiltonian extensions, whose recursion relations are in a case of  ``unsplit'' and in the other of ``split'' type. This suggests the possibility to deal with both cases in a unified way.

The paper is organized in the following way: in the next section the four separable set of coordinates on the Euclidean plane are briefly presented, as well as the pair of quadratic Hamiltonians associated with each coordinate systems and the corresponding St\"ackel matrix. Because the Euclidean plane is a constant curvature manifold, separation of variables only occurs in orthogonal coordinates and therefore the presented list is exhaustive.

Section three first presents the procedure for constructing the extended systems in the two asymmetric sets of coordinates, moreover some basic concepts of the bihamiltonian and trihamiltonian framework are reviewed. In the following the procedure is separately adapted to each of the two symmetric coordinate systems; this allows to show how this method could be generalized to other cases.

Finally, some explicit examples are shown. In particular the one-Casimir extensions of the H\'enon--Heiles and Kepler systems presented in \cite{Bla2000} are recovered from a trihamiltonian point of view and it is shown, in the Kepler case, how multiple sets of separation coordinates lead to different trihamiltonian extension of the same Hamiltonian.

\section{Systems related to separable orthogonal web in the plane}
It is well known that, in the Euclidean plane, there are only four systems of orthogonal coordinates allowing to separate the Hamilton-Jacobi equation associated to a natural Hamiltonian \cite{BenGen}. As a matter of fact, instead of considering the coordinate system, it is often appropriate to consider the associated web, i.e. the family of curves on which the coordinates are constant because the web associated to a coordinate systems is invariant respect to transformation $q_i\mapsto Q_i$ such that the new coordinate $Q_i$ turns out to depend only on $q_i$ for all fixed $i$. This kind of transformation of coordinates is called \emph{separated}. If $x$ and $y$ denote the Cartesian coordinates in the Euclidean plane, then the other three coordinate systems are:
\begin{itemize}
\item parabolic coordinates:
\BD
\begin{array}{ll}
x = \frac{u+v}{2} & u = x + \sqrt{x^2+y^2} \\[5pt]
y = \sqrt{-uv} & v = x - \sqrt{x^2+y^2}
\end{array}
\ED
whose associated web is made up of confocal parabolae with focus in the origin and symmetric respect to the x-axis;
\item elliptic-hyperbolic coordinates:
\BD
\begin{array}{ll}
x = \frac{ds}{2k} & s = \sqrt{(x+k/2)^2+y^2} + \sqrt{(x-k/2)^2+y^2} \\[5pt]
y = \frac{\sqrt{-(d^2-k^2)(s^2-k^2)}}{2k} & d = \sqrt{(x+k/2)^2+y^2} - \sqrt{(x-k/2)^2+y^2}
\end{array}
\ED
whose web is made up of a family of confocal hyperbolae and a family of confocal ellipses, whose common foci lie on the x-axis with coordinate~$\pm k/2$;
\item polar coordinates:
\BD
\begin{array}{ll}
x = r\,\cos\theta & r = \sqrt{x^2+y^2} \\[5pt]
y = r\,\sin\theta & \theta = \arctan(y/x)
\end{array}
\ED
centred at the origin.
\end{itemize}

The study of the Killing tensors associated to each coordinate system allows to construct four canonical types of completely integrable systems (see \cite{Cha2001} and references therein). In fact, a pair of Hamiltonian functions $H$ and $K$, in involution with respect to the canonical Poisson bracket on the cotangent bundle of $\Reali^2$, and both separable in the given coordinates, is related to each coordinate system (more precisely to each web underlying the coordinate system). The separability of these Hamiltonians is ensured by the following relations involving the St\"ackel matrix of the coordinate system.
\begin{description}
\item{\textbf{Cartesian coordinates}}\\
Hamiltonian functions:
\begin{eqnarray*}
H &=& \half p_x^2 + \phi_1(x) \\
K &=& \half p_y^2 + \phi_2(y) .
\end{eqnarray*}
St\"ackel relation:
\BD
\left(
\begin{array}{cc}
1 & 0 \\[5pt]
0 & 1
\end{array}
\right)
\left(
\begin{array}{c}
H \\[5pt]
K
\end{array}
\right)
=
\left(
\begin{array}{c}
\half p_x^2 + \phi_1(x) \\[5pt]
\half p_y^2 + \phi_2(y)
\end{array}
\right).
\ED
\item{\textbf{Parabolic coordinates}}\\
Hamiltonian functions:
\begin{eqnarray*}
H &=& \frac{\left(\half p_u^2+\phi_1(u)\right)u
-\left(\half p_v^2+\phi_2(v)\right)v}{u-v} \\
K &=& \frac{\left(\half p_v^2-\half p_u^2 +\phi_2(v)-\phi_1(u)\right)uv}{u-v}.
\end{eqnarray*}
St\"ackel relation:
\BD
\left(
\begin{array}{cc}
1 & \frac{1}{u}\\[5pt]
1 & \frac{1}{v}
\end{array}
\right)
\left(
\begin{array}{c}
H \\[5pt]
K
\end{array}
\right)
=
\left(
\begin{array}{c}
\half p_u^2+\phi_1(u) \\[5pt]
\half p_v^2+\phi_2(v)
\end{array}
\right).
\ED
\item{\textbf{Elliptic-hyperbolic coordinates}}\\
Hamiltonian functions:
\begin{eqnarray*}
H &=& \frac{\left(\half p_s^2+\phi_1(s)\right)(s^2-k^2)
-\left(\half p_d^2+\phi_2(d)\right)(d^2-k^2)}{s^2-d^2} \\
K &=& \frac{\left(\half p_d^2-\half p_s^2 +\phi_2(d)-\phi_1(s)\right)(s^2-k^2)(d^2-k^2)}{s^2-d^2}.
\end{eqnarray*}
St\"ackel relation:
\BD
\left(
\begin{array}{cc}
1 & \frac{1}{s^2-k^2}\\[5pt]
1 & \frac{1}{d^2-k^2}
\end{array}
\right)
\left(
\begin{array}{c}
H \\[5pt]
K
\end{array}
\right)
=
\left(
\begin{array}{c}
\half p_s^2+\phi_1(s) \\[5pt]
\half p_d^2+\phi_2(d)
\end{array}
\right).
\ED
\item{\textbf{Polar coordinates}}\\
Hamiltonian functions:
\begin{eqnarray*}
H &=& \half p_r^2+\phi_1(r)+\frac{1}{r^2}\left(\half p_\theta^2+\phi_2(\theta)\right)\\
K &=& -\half p_\theta^2-\phi_2(\theta).
\end{eqnarray*}
St\"ackel relation:
\BD
\left(
\begin{array}{cc}
1 & \frac{1}{r^2} \\[5pt]
0 & 1
\end{array}
\right)
\left(
\begin{array}{c}
H \\[5pt]
K
\end{array}
\right)
=
\left(
\begin{array}{c}
\half p_r^2+\phi_1(r) \\[5pt]
-\half p_\theta^2-\phi_2(\theta)
\end{array}
\right).
\ED
\end{description}
Before going ahead a remark on the choice of the Hamiltonians is needed: sometimes, in Cartesian coordinates, the two Hamiltonian functions
\BAS
\bar{H}&=&\half(p_x^2+p_y^2)+\phi_1(x)+\phi_2(y)\\
\bar{K}&=&\half p_y^2+\phi_2(y)
\EAS
are considered, with the St\"ackel matrix
$$
S=\left(
\begin{array}{cc}
1 & -1 \\
0 & 1
\end{array}
\right)
$$

The Hamiltonians $H$ and $K$ used in the present work are trivial recombinations of Hamiltonians $\bar{H}$ and $\bar{K}$ that allow to put the St\"ackel matrix in a more useful form. This form is such that in the $i$-th row there is just an ordered sequence (possibly decreasing) of powers of a suitable function of the coordinate $q_i$. One of the key point in order to extend the procedure presented in this paper to more general cases is to characterize the class of St\"ackel matrix that can be put in this form.

It is worth to observe that, with an appropriate change of variables, the previous four types of separable systems are mapped into the four types of integrable systems in the plane with two quadratic first integrals (see, for example \cite{Per1990}). The method to construct a trihamiltonian extension of the previous systems is slightly different for the asymmetric coordinate systems (parabolic and elliptic-hyperbolic) and for the symmetric ones (polar and Cartesian). This difference is due to the different form of the St\"ackel matrix in the two cases, as it can be seen from the above formulae. In fact, for asymmetric coordinate systems all the rows of the St\"ackel matrix have the same form and differ only in the current coordinate, whereas for symmetric coordinate systems different rows of the St\"ackel matrix have different forms. Therefore, the method that will be used in the asymmetric case will need to be adapted to each symmetric coordinate systems in a specific way.

\section{Trihamiltonian extension of separable systems}
\subsection{Asymmetric coordinate systems}
The idea of constructing a trihamiltonian extension of separable systems originates from the
construction of bihamiltonian extension of Benenti systems presented in \cite{Mag2000}, and it is
motivated by some results about trihamiltonian systems presented in \cite{mio} that will be briefly summarized. If $\Pp$, $\Q$ and $\R$ are three Poisson structures mutually compatible on a manifold of dimension $2n+k$, and $f$ is simultaneously a Casimir function for the two Poisson pencils $\Q-\lambda\Pp$ and $\R-\mu\Pp$, polynomial in $\lambda$ and $\mu$, then all the coefficients $h_k$ of this polynomial are functions mutually in involution with respect to all the three Poisson structures, and satisfy a recurrence scheme depending on the form of the polynomial $f$. Under some further hypotheses, it is possible to find out a set of coordinates $\{\lambda_i,\mu_i, c_\alpha\}$ with the following characteristics:
\begin{enumerate}
\item $c_\alpha$ are $k$ Casimir functions for $\Pp$
\item $\lambda_i,\mu_i$ are $2n$ Darboux--Nijenhuis coordinates for both the Nijenhuis tensors obtained, thanks to a deformation procedure, from the two Poisson pencils $\Q-\lambda\Pp$ e $\R-\mu\Pp$
\item $\lambda_i,\mu_i$ satisfy the $n$ relations:
\BE\label{separo}
f(\lambda_i,\mu_i;\{h_k\},{c_\alpha}) = r_i(\lambda_i,\mu_i)
\EE
in which the functions $r_i(\lambda,\mu)$ are constant functions on the manifold.
\end{enumerate}
The relations (\ref{separo}) imply the separability (in Sklyanin sense) of Hamiltonians~$h_k$. It is then natural to ask if it is possible, starting from a classical separable system, to obtain an adapted trihamiltonian structure. 

As a matter of fact, the separation relations encoded by the St\"ackel matrix in asymmetric coordinate systems are very similar to relations (\ref{separo}), in the special case in which the dimension of the symplectic leaves of $\Pp$ is $n=2$. Indeed with the symplectic (and separated) transformations:
\BD
\lambda_1 = u, \lambda_2 = v, \mu_1 = p_u, \mu_2 = p_v
\ED
for parabolic coordinates and
\BD
\lambda_1 = s^2-k^2, \lambda_2 = d^2-k^2, \mu_1 = \frac{p_s}{2s}, \mu_2 = \frac{p_d}{2d}
\ED
for elliptic--hyperbolic coordinates, the relations extracted from St\"ackel matrix become:
$$
\left(
\begin{array}{cc}
\lambda_1 & 1 \\
\lambda_2 & 1
\end{array}
\right)
\left(
\begin{array}{c}
H \\
K
\end{array}
\right)=
\left(
\begin{array}{c}
r_1(\lambda_1,\mu_1) \\
r_2(\lambda_1,\mu_1)
\end{array}
\right)
$$
that is
\BE\label{nucleo_asim}
K + \lambda_i H=r_i(\lambda_i,\mu_i)
\EE
where the functions $r_i(\lambda,\mu)$ are, respectively in the two coordinate systems:
\begin{eqnarray*}
r_i(\lambda,\mu) &=& \half\lambda\mu^2+\lambda\phi_i(\lambda) \\
r_i(\lambda,\mu) &=& 2\mu^2(\lambda^2+\lambda k^2)+\lambda\phi_i(\sqrt{\lambda+k^2})\,.
\end{eqnarray*}

The difference between relations (\ref{separo}) and (\ref{nucleo_asim}) is the absence in the second ones of any Casimir function $c_\alpha$; this implies that the polynomial $K+\lambda H$ can not be a Casimir function for any Poisson pencil, including the canonical Poisson structure on the cotangent bundle of $\Reali^2$. In order to obtain a set of relations analogous to (\ref{separo}) it is necessary to extend the phase space of the system with a suitable number of coordinates $c_\alpha$, and to define on the extended space a Poisson structure $\Pp$ such that the new coordinates are its Casimir functions, and the reduction of $\Pp$ on its symplectic leaf $\{c_\alpha=0\}$ gives the original Poisson bracket. This allows to think relations (\ref{nucleo_asim}) as a reduction of relations (\ref{separo}) on the symplectic leaf $\{c_\alpha=0\}$. 
Moreover, in the extended space, the functions that reduce to $H$ and $K$ could depend on the new coordinates $c_\alpha$, hence it is also necessary to introduce two ``deformed'' Hamiltonians $\widetilde{H}$ and $\widetilde{K}$. Taking in account these observations the simplest function analogous to the one appearing in (\ref{separo}) and compatible with the relations (\ref{nucleo_asim}) is:
\BE\label{cas_asim}
f=\widetilde{K}+\lambda\widetilde{H}+c_1\lambda^2+\mu c_2+\lambda\mu c_3
\EE
where $\widetilde{H}$ and $\widetilde{K}$ are the appropriate extensions of the Hamiltonians $H$ and $K$ and three extra coordinates are introduced.
The extended Poisson structure $\Pp$ has hence the form
\BD
\Pp=\left(
\begin{array}{ccccccc}
0&0&1&0&0&0&0 \\
0&0&0&1&0&0&0\\
-1&0&0&0&0&0&0\\
0&-1&0&0&0&0&0\\
0&0&0&0&0&0&0\\
0&0&0&0&0&0&0\\
0&0&0&0&0&0&0
\end{array}
\right)
\ED
And the trihamiltonian recursion scheme associated to the function (\ref{cas_asim}) is:
\BE\label{ric_asim}
\begin{picture}(140,80)
\put(0,0){\Scatola{$\widetilde{K}$}}
\put(15,15){\Nucleo{$X_{\widetilde{K}}$}}
\put(30,60){\Scatola{$c_2$}}
\put(60,0){\Scatola{$\widetilde{H}$}}
\put(75,15){\Nucleo{$X_{\widetilde{H}}$}}
\put(90,60){\Scatola{$c_3$}}
\put(120,0){\Scatola{$c_1$}}
\end{picture}
\EE
In the previous scheme the notation $f\stackrel{P}{\longrightarrow}X_f$ means that the vector field $X_f$ is obtained from the differential of the function $f$ through the Poisson structure $P$. Further, if the relation $P\D f=Q\D g$ between the two functions $f$ and $g$ holds, it is sometimes said (understanding some privileged role for the Poisson structure $P$) that $f$ is the \emph{bihamiltonian antecedent} of $f$ through the structure $Q$.

The previous construction needs some remarks: first, it is possible to reduce the number of new coordinates that is needed by omitting the ``vertical'' part of the recurrence scheme and looking for a function in the form:
\BD
f=\widetilde{K}+\lambda\widetilde{H}+\lambda^2 c_1 \,.
\ED
This approach is followed in \cite{Mag2000} and \cite{Bla2000}, but allows to obtain just  a bihamiltonian recursion. On the other hand it is  possible, both in the bihamiltonian (as in \cite{Bla2000}) and in the trihamiltonian case, to add to the function (\ref{cas_asim}) a ``redundant'' Casimir function, that is a Casimir function common to all the Poisson structures, using the function:
\BD
\bar{f}=f+\lambda^n c_4 \mbox{ with } n > 2\,.
\ED
In this way a different pair of extended Hamiltonians $\widetilde{H}$ and $\widetilde{K}$ is constructed and it can be reduced to that previously obtained putting $c_4=0$. Even the new Poisson structures can be related to the old ones through the reduction on the zero level set of the
common Casimir function $c_4$.

Besides these digressions, the next step of the procedure is to require that the coordinates $\lambda_i,\mu_i$ are separation
coordinates in the sense of (\ref{separo}). This implies that $\widetilde{H}$ and $\widetilde{K}$ must  solve the linear system
$$
f(\lambda_i,\mu_i;\widetilde{K},\widetilde{H},c_\alpha)=r_i(\lambda_i,\mu_i)
$$
Therefore the solution, using the function (\ref{cas_asim}), is:
\begin{eqnarray*}
\widetilde{H} &=& \frac{r_2(\lambda_2,\mu_2)-r_1(\lambda_1,\mu_1)}{\lambda_2-\lambda_1} - \\
&& (\lambda_1+\lambda_2)c_1 +
\frac{\left (\mu_2-\mu_1\right)}{\lambda_1-\lambda_2}c_2 + 
\frac{\left(\mu_2\lambda_2-\mu_1\lambda_1\right )}{\lambda_1-\lambda_2}c_3\,; \\ 
\widetilde{K} &=& \frac{\lambda_2 r_1(\lambda_1,\mu_1) - \lambda_1
r_2(\lambda_2,\mu_2)}{\lambda_2-\lambda_1} + \\ && \lambda_2\lambda_1c_1
+\frac{\left(\lambda_2\mu_1-\mu_2\lambda_1\right)}{\lambda_1-\lambda_2}c_2 +
\frac{\lambda_2\lambda_1\left (\mu_2-\mu_1\right )}{\lambda_2-\lambda_1}c_3\,.
\end{eqnarray*}
The last step, in order to obtain the desired recursion scheme,  is to construct other two Poisson
structures, after the $\Pp$ defined above. Following \cite{mio}
one start defining the two ``deformed'' Poisson tensors represented by the $7\times 7$ matrices
\begin{eqnarray*}
\Qd &=& \left(
\begin{array}{ccc}
0 & \Lambda & 0 \\
-\Lambda & 0 & 0 \\
0 & 0 & 0
\end{array}
\right) \\
\Rd &=& \left(
\begin{array}{ccc}
0 & \Mu & 0 \\
-\Mu & 0 & 0 \\
0 & 0 & 0
\end{array}
\right)
\end{eqnarray*}
where
\begin{eqnarray*}
\Lambda &=& \left(
\begin{array}{cc}
\lambda_1 & 0 \\
0 & \lambda_2
\end{array}
\right) \\
\Mu &=& \left(
\begin{array}{cc}
\mu_1 & 0 \\
0 & \mu_2
\end{array}
\right)
\end{eqnarray*}
and then the two vector fields by constructing:
\begin{eqnarray*}
\Xq &=& \textstyle{\sum_\alpha} F_\alpha \frac{\partial}{\partial c_\alpha} \\
\Xr &=& \textstyle{\sum_\alpha} G_\alpha \frac{\partial}{\partial c_\alpha}
\end{eqnarray*}
where the functions $F_\alpha$ and $G_\alpha$ are the bihamiltonian antecedents of the Casimir functions of
$\Pp$ through, respectively the tensors $\Q$ e $\R$. This means that relatively to the recursion
scheme (\ref{ric_asim}), they satisfy the relations $\Pp\D F_\alpha = \Q\D c_\alpha$ e $\Pp\D
G_\alpha =
\R\D c_\alpha$. Hence in the given case it holds $F_1=\widetilde{H},\; F_2=F_3=0$ and $G_1=0,\;
F_2=\widetilde{K},\;F_3=\widetilde{H}$. The two vector fields then are:
\begin{eqnarray*}
\Xq &=& \widetilde{H} \frac{\partial}{\partial c_1} \\
\Xr &=& \widetilde{K} \frac{\partial}{\partial c_2} + \widetilde{H} \frac{\partial}{\partial c_3}.
\end{eqnarray*}
Lastly, it is possible to construct the two tensors
\begin{eqnarray*}
\Q &=& \Qd - \DLie{\Xq}{\Pp}  \\
\R &=& \Rd - \DLie{\Xr}{\Pp}
\end{eqnarray*}
The surprising result of this complicated construction is the following fact, that could be verified
by direct computation:
\begin{Fatto}
The tensors $\Q$ and $\R$ previously constructed are compatible Poisson tensors independently of
the choice of the functions $r_i(\lambda,\mu)$. Moreover, the function $f$ given by (\ref{cas_asim})
is a common Casimir function for the two Poisson pencils $\Q-\lambda\Pp$ and $\R-\lambda\Pp$,
realizing the recursion scheme (\ref{ric_asim}).
\end{Fatto}

Being $c_2$ and $c_3$ two common Casimir functions of $\Pp$ and $\Q$, both these Poisson structure are reducible on the level surface $\{c_2=c_3=0\}$. On this surface the function (\ref{cas_asim}) reduce itself to
$$
\widetilde{K}+\lambda\widetilde{H}+\lambda^2 c_1
$$
and the trihamiltonian recursion scheme (\ref{ric_asim}) becomes a simply bihamiltonian one. Because the separation relation (\ref{separo}) for the reduced systems becomes
\BAS
\widetilde{K}+\lambda_1\widetilde{H}+\lambda_1^2 c_1 &=& r_1(\lambda_1,\mu_1) \\
\widetilde{K}+\lambda_2\widetilde{H}+\lambda_2^2 c_1 &=& r_2(\lambda_2,\mu_2)
\EAS
when $\phi_1=\phi_2$, and then $r_1=r_2=r(\lambda,\mu)$, the function
$$
\widetilde{K}+\lambda\widetilde{H}+\lambda^2 c_1-r(\lambda,\mu)
$$
define an unsplitted separation curve in the sense of \cite{Bla2000}.

On the contrary, reducing the function $f$ and the Poisson structures $\Pp$ and $\R$ on the level surface $\{c_1=0\}$ of their common Casimir function $c_1$, a splitted bihamiltonian recursion chain is obtained.

\subsection{Symmetric coordinate systems}
The procedure described above cannot be directly applied to Hamiltonians separable in a symmetric coordinate system, i.e. in polar or Cartesian coordinates. In these cases, actually, the relationships obtained from the St\"ackel matrix cannot be summarized with an unique polynomial, able to suggest a  candidate $f$ to the role of  common Casimir function to the two Poisson pencils $\Q-\lambda\Pp$ and $\R-\lambda\Pp$. This difficulty could be overcame observing that a trihamiltonian structure doesn't admit necessarily only one common Casimir function for the two Poisson pencils. Therefore, it will be sufficient to build a different common Casimir function for each of the different polynomial relations generated by the St\"ackel matrix. 

In the case of polar coordinates, through the symplectic transformations:
\BD
\lambda_1 = r^2, \lambda_2 = \tan\theta, \mu_1 = \frac{p_r}{2r}, \mu_2 = \cos^2\theta\, p_\theta 
\ED
the St\"ackel relationships could be rewritten with the two equations:
\BE\label{nucleo_pol}
\begin{array}{rcl}
K + \lambda_1 H &=& r_1(\lambda_1,\mu_1) \\
K &=& r_2(\lambda_2,\mu_2)
\end{array}
\EE
where the functions $r_i(\lambda,\mu)$ are:
\begin{eqnarray*}
r_1(\lambda,\mu) &=& 2\mu^2\lambda^2+\lambda\phi_1(\sqrt{\lambda}) \\
r_2(\lambda,\mu) &=&
-\half(1+\lambda^2)^2\mu^2-\phi_2(\arctan(\lambda))
\end{eqnarray*}
Hence two functions have to be constructed, with the form:
\BE\label{cas_pol}
\begin{array}{rcl}
f_1 &=& \widetilde{K}+\lambda\widetilde{H}+\lambda^2c_1+\mu c_2+\lambda\mu c_3 \\
f_2 &=& \widetilde{K}+\lambda c_4 + \mu c_2
\end{array}
\EE
and realizing the trihamiltonian recursion scheme:
\BE\label{ric_pol}
\begin{picture}(240,80)
\put(0,0){\Scatola{$\widetilde{K}$}}
\put(15,15){\Nucleo{$X_{\widetilde{K}}$}}
\put(30,60){\Scatola{$c_2$}}
\put(60,0){\Scatola{$\widetilde{H}$}}
\put(75,15){\Nucleo{$X_{\widetilde{H}}$}}
\put(90,60){\Scatola{$c_3$}}
\put(120,0){\Scatola{$c_1$}}
\put(160,0){\Scatola{$\widetilde{K}$}}
\put(175,15){\Nucleo{$X_{\widetilde{K}}$}}
\put(190,60){\Scatola{$c_2$}}
\put(220,0){\Scatola{$c_4$}}
\end{picture}
\EE
The expressions for $\widetilde{H}$ and $\widetilde{K}$ can be obtained by solving the system:
\BD
\left\{
\begin{array}{l}
\widetilde{K}+\lambda_1\widetilde{H}+\lambda_1^2c_1+\mu_1 c_2+\lambda_1\mu_1 c_3 = r_1(\lambda_1,\mu_1) \\
\widetilde{K}+\lambda_2 c_4 + \mu_2 c_2  = r_2(\lambda_2,\mu_2)
\end{array}
\right.
\ED
whose solution is:
\begin{eqnarray*}
\widetilde{H} &=& \frac{r_1(\lambda_1,\mu_1)-r_2(\lambda_2,\mu_2)}{\lambda_1}
-\lambda_1 c_1 +\frac{\mu_2-\mu_1}{\lambda_1}c_2 -\mu_1 c_3+\frac{\lambda_2}{\lambda_1}c_4 \\
\widetilde{K} &=& r_2(\lambda_2,\mu_2)-\mu_2 c_2-\lambda_2c_4
\end{eqnarray*}
The $\Pp$, $\Q$ and $\R$ structures are obtained, analogously to the asymmetric case, by adding to the tensors $\Qd$ and $\Rd$ the Lie derivative of $\Pp$ respect to the vector fields $\Xq$ and $\Xr$, constructed with the antecedents of the four Casimir functions of $\Pp$. From the recursion scheme (\ref{ric_pol}) these two vector fields turn out to be:
\begin{eqnarray*}
\Xq &=& \widetilde{H}\frac{\partial}{\partial c_1} + \widetilde{K}\frac{\partial}{\partial c_4} \\
\Xr &=& \widetilde{K}\frac{\partial}{\partial c_2} + \widetilde{H}\frac{\partial}{\partial c_3}.
\end{eqnarray*}
By direct calculation the following fact can be proved:
\begin{Fatto}
The tensors $\Q$ and $\R$ previously constructed are compatible Poisson tensors, independently of
the choice of the functions $r_i(\lambda,\mu)$. Moreover, the functions $f_1$ and $f_2$ given by
(\ref{cas_pol}) are two common Casimir functions for the two Poisson pencils $\Q-\lambda\Pp$ and
$\R-\lambda\Pp$, thus realizing the recursion scheme (\ref{ric_pol}).
\end{Fatto}

Eventually, in the case of Cartesian coordinates, the starting point is the construction of two functions of the form:
\BE\label{cas_car}
\begin{array}{rcl}
f_1 &=& \widetilde{H}+c_1\lambda+\mu c_2 \\
f_2 &=& \widetilde{K}+\lambda c_3 + \mu c_4
\end{array}
\EE
that realize the trihamiltonian recursion scheme:
\BE\label{ric_car}
\begin{picture}(240,80)
\put(0,0){\Scatola{$\widetilde{H}$}}
\put(15,15){\Nucleo{$X_{\widetilde{H}}$}}
\put(30,60){\Scatola{$c_2$}}
\put(60,0){\Scatola{$c_1$}}
\put(100,0){\Scatola{$\widetilde{K}$}}
\put(115,15){\Nucleo{$X_{\widetilde{K}}$}}
\put(130,60){\Scatola{$c_4$}}
\put(160,0){\Scatola{$c_3$}}
\end{picture}
\EE
By solving the linear system:
\BD
\left\{
\begin{array}{l}
\widetilde{H}+c_1\lambda_1+\mu_1 c_2 = r_1(\lambda_1,\mu_1) \\
\widetilde{K}+\lambda_2 c_3 + \mu_2 c_4  = r_2(\lambda_2,\mu_2)
\end{array}
\right.
\ED
it is possible to find out the Hamiltonians:
\begin{eqnarray*}
\widetilde{H}  &=& r_1(\lambda_1,\mu_1) -c_1\lambda_1-\mu_1 c_2 \\
\widetilde{K}  &=& r_2(\lambda_2,\mu_2) -\lambda_2 c_3 - \mu_2 c_4
\end{eqnarray*}
and from the recursion scheme (\ref{ric_car}) the following vector fields are obtained:
\begin{eqnarray*}
\Xq &=& \widetilde{H}\frac{\partial}{\partial c_1} + \widetilde{K}\frac{\partial}{\partial c_3} \\
\Xr &=& \widetilde{H}\frac{\partial}{\partial c_2} + \widetilde{K}\frac{\partial}{\partial c_4}.
\end{eqnarray*}
Even in this case it holds:
\begin{Fatto}
The tensors $\Q$ and $\R$ previously constructed are compatible Poisson tensors, independently of
the choice of the functions $r_i(\lambda,\mu)$. Moreover, the functions $f_1$ and $f_2$ given by
(\ref{cas_pol}) are two common Casimir functions for the two Poisson pencils $\Q-\lambda\Pp$ and
$\R-\lambda\Pp$, thus realizing the recursion scheme (\ref{ric_car}).
\end{Fatto}

\section{Examples}
\subsection{Trihamiltonian extension of H\'enon--Heiles system}
An application of the recipe presented above is given by the H\'enon--Heiles system, already considered in \cite{Bla2000}, whose Hamiltonian expressed in natural coordinates $\{q_1,q_2,p_1,p_2\}$ is:
\BD
H=\half(p_1^2+p_2^2) + \half q_1q_2^2 + q_1^3
\ED
This system is separable in parabolic coordinates and the previous Hamiltonian can be obtained from the general expression through the coordinate transformation:
\BD
\begin{array}{ll}
q_1 = \lambda_1+\lambda_2, &
q_2 = 2\sqrt{-\lambda_1\lambda_2} \\
p_1 = \frac{\lambda_1\mu_1-\lambda_2\mu_2}{\lambda_1-\lambda_2}, &
p_2 = \sqrt{-\lambda_1\lambda_2}\frac{\mu_1-\mu_2}{\lambda_1-\lambda_2}
\end{array}
\ED
and choosing the two arbitrary functions $\phi_1$ e $\phi_2$ to be:
\BD
\phi_1(z)=\phi_2(z)=z^3
\ED
Applying the explained procedure one get the two deformed Hamiltonians
\begin{eqnarray*}
\widetilde{H} &=& \half(p_1^2+p_2^2)+\half q_1q_2^2+q_1^3-q_1c_1-2\frac{p_2}{q_2}c_2-p_1c_3 \\
\widetilde{K} &=& \half(q_2p_1p_2-q_1p_2^2)+\frac{1}{4}q_1^2q_2^2+\frac{1}{16}q_2^4
-\frac{1}{4}q_2^2c_1+(2\frac{q_1p_2}{q_2}-p_1)c_2-\half q_2p_2c_3
\end{eqnarray*}
that are put in the recursion scheme (\ref{ric_asim}) by the three Poisson structures:
\BD
\Pp=\left(
\begin{array}{ccccccc}
0&0&1&0&0&0&0 \\
0&0&0&1&0&0&0\\
-1&0&0&0&0&0&0\\
0&-1&0&0&0&0&0\\
0&0&0&0&0&0&0\\
0&0&0&0&0&0&0\\
0&0&0&0&0&0&0
\end{array}
\right)
\ED
\BD
\Q=\left(
\tiny{\begin{array}{ccccccc}
0&0&q_1&\half q_2&p_1-c_3&0&0\\
0&0&\half q_2&0&p_2-2 \frac{c_2}{q_2}&0&0\\
-q_1&-\half q_2&0&\half p_2&c_1-3 q_1^2-\half q_2^2&0&0\\
-\half q_2&0&-\half p_2&0&-q_1q_2-2 \frac{p_2}{q_2^2}c_2&0&0\\
c_3-p_1&2 \frac{c_2}{q_2}-p_2&3 q_1^2+\half q_2^2-c_1&q_1q_2+2 \frac{p_2}{q_2^2}c_2
&0&0&0\\
0&0&0&0&0&0&0\\
0&0&0&0&0&0&0
\end {array}}
\right)
\ED
\BD
\R=\left(
\tiny{\begin{array}{ccccccc}
0&0&p_1&p_2&0&Y_1&X_1\\
0&0&p_2&p_1-2\frac{p_2q_1}{q_2}&0&Y_2&X_2\\
-p_1&-p_2&0&\frac{p_2^2}{q_2}&0&Y_3&X_3\\
-p_2& 2\frac{q_1p_2}{q_2}-p_1&-\frac{p_2^2}{q_2}&0&0&Y_4&X_4\\
0&0&0&0&0&0&0\\
-Y_1&-Y_2&-Y_3&-Y_4&0&0&0\\
-X_1&-X_2&-X_3&-X_4&0&0&0
\end{array}}
\right)
\ED
where $X=X_{\widetilde{H}}=\Pp\D\widetilde{H}$ and $Y=X_{\widetilde{K}}=\Pp\D\widetilde{K}$.
It is worth to observe that both $\Pp$ and $\Q$ can be reduced by restriction on the level surface $c_2=c_3=0$ and that the reduced Hamiltonians and Poisson structures are the ones considered in \cite{Bla2000}. Instead the structure $\R$ isn't reducible, so the reduced system appear to be just a one--Casimir bihamiltonian extension of the H\'enon--Heiles system. An advantage of the knowledge of a trihamiltonian extension of the system is the possibility to perform a reduction on the level set
$c_1=0$, producing a different, two--Casimir bihamiltonian extension of the system, whose Poisson structures are the reduction of $\Pp$ and $\R$.

\subsection{Trihamiltonian extension of Kepler system in the plane}
The Kepler system in the plane is separable in three different set of coordinates: the parabolic, the elliptic-hyperbolic and the polar coordinate systems. To each of these sets of coordinates can be associated a different second constant of motion (related to the Killing tensor of the coordinate system) and a different trihamiltonian extension.

In the case of parabolic coordinates the transformation from the natural coordinates to the coordinates $\{\lambda_i,\mu_i\}$ is given by:
\BD
\begin{array}{ll}
q_1 = 2\sqrt{-\lambda_1\lambda_2}, &
q_2 = \lambda_1+\lambda_2 \\
p_1 = \frac{\sqrt{-\lambda_1\lambda_2}(\mu_1-\mu_2)}{\lambda_1-\lambda_2}, &
p_2 = \frac{\lambda_1\mu_1-\lambda_2\mu_2}{\lambda_1-\lambda_2}
\end{array}
\ED
and setting
\begin{eqnarray*}
\phi_1(z) &=& \frac{a}{2z}, \\
\phi_2(z) &=& -\frac{a}{2z}
\end{eqnarray*}
the deformed Hamiltonians $\widetilde{H}$ and $\widetilde{K}$ become
\begin{eqnarray*}
\widetilde{H} &=&
\half(p_1^2+p_2^2)-\frac{a}{\sqrt{q_1^2+q_2^2}}
-q_2c_1-2\frac{p_1}{q_1}c_2-p_2c_3 \\
\widetilde{K} &=&
\half(q_1p_1p_2-q_2p_1^2)+\frac{aq_2}{2\sqrt{q_1^2+q_2^2}}
-\frac{1}{4}q_1^2c_1+\frac{2q_2p_1-q_1p_2}{q_1}c_2-\half q_1p_1c_3\,.
\end{eqnarray*}
They are put in the recursion scheme (\ref{ric_asim}) by the three Poisson structures
\BD
\Pp=\left(
\begin{array}{ccccccc}
0&0&1&0&0&0&0 \\
0&0&0&1&0&0&0\\
-1&0&0&0&0&0&0\\
0&-1&0&0&0&0&0\\
0&0&0&0&0&0&0\\
0&0&0&0&0&0&0\\
0&0&0&0&0&0&0
\end{array}
\right)
\ED
\BD
\Q = \left(
\scriptsize{\begin{array}{ccccccc}
0&0&0&\half q_1&p_1-2 \frac{c_2}{q_1} &0&0\\
0&0&\half q_1&q_2&p_2-c_3&0&0\\
0&-\half q_1&0&-\half p_1&-2\frac{p_1}{q_1^2}c_2-\frac{aq_1}{\left(q_1^2+q_2^2\right)^\frac{3}{2}}&0&0\\
-\half q_1&-q_2&\half p_1&0&c_1-\frac{aq_2}{\left(q_1^2+q_2^2\right)^\frac{3}{2}}&0&0\\
2 \frac{c_2}{q_1}-p_1 &c_3-p_2&2\frac{p_1}{q_1^2}c_2+\frac{aq_1}{\left(q_1^2+q_2^2\right)^\frac{3}{2}}&\frac{aq_2}{\left(q_1^2+q_2^2\right)^\frac{3}{2}}-c_1&0&0&0\\
0&0&0&0&0&0&0\\
0&0&0&0&0&0&0
\end{array}}
\right)
\ED
\BD
\R = \left(
\small{\begin{array}{ccccccc}
0&0&p_2-2 \frac{p_1q_2}{q_1}&p_1&0&Y_1&X_1\\
0&0&p_1&p_2&0&Y_2&X_2\\
2\frac{p_1q_2}{q_1}-p_2&-p_1&0&
-\frac{p_1^2}{q_1}&0&Y_3&X_3\\
-p_1&-p_2&\frac{p_1^2}{q_1}&0&0&Y_4&X_4\\
0&0&0&0&0&0&0\\
-Y_1&-Y_2&-Y_3&-Y_4&0&0&0\\
-X_1&-X_2&-X_3&-X_4&0&0&0
\end{array}}
\right)
\ED
where $X=X_{\widetilde{H}}=\Pp\D\widetilde{H}$ and $Y=X_{\widetilde{K}}=\Pp\D\widetilde{K}$.
As can be easily seen, reducing the Poisson structures $\Pp$ and $\Q$ and the Hamiltonians on the level set $c_2=c_3=0$, it is obtained the one-Casimir extension considered in \cite{Bla2000}. But it is important to observe that the separation coordinates used don't seem to originate a unique ``separation curve'' because the two
functions $\phi_1$ and $\phi_2$ are different.

In the case of the elliptic-hyperbolic coordinates the coordinates transformation from the natural coordinates is:
\BD
\begin{array}{ll}
q_2 = \frac{\sqrt{-\lambda_1\lambda_2}}{k}, &
q_1 = \frac{\sqrt{(\lambda_1+k^2)(\lambda_2+k^2)}}{k}+k \\
p_1 =2\frac{(\lambda_1\mu_1-\lambda_2\mu_2)\sqrt{(\lambda_1+k^2)(\lambda_2+k^2)}}
{k(\lambda_1-\lambda_2)},
& p_2 =
2\frac{\mu_1(\lambda_1+k^2)-\mu_2(\lambda_2+k^2)}{k(\lambda_1-\lambda_2)}
\sqrt{-\lambda_1\lambda_2}
\end{array}
\ED
and the two arbitrary functions are
\BD
\phi_1(z) = \phi_2(z) = \frac{az}{k^2-z^2}\,.
\ED
The two deformed Hamiltonians $\widetilde{H}$ and $\widetilde{K}$ turn out to be:
\begin{eqnarray*}
\widetilde{H} &=& \half(p_1^2+p_2^2)-\frac{a}{\sqrt{q_1^2+q_2^2}} \\
&&-(q_1^2+q_2^2-2q_1k)c_1
+\frac{1}{2k^2}\left(\frac{p_1}{(q_1-k)}-\frac{p_2}{q_2}\right)c_2
-\frac{p_1}{2(q_1-k)}c_3
\\
\widetilde{K} &=&
(q_1-k)q_2p_1p_2-\half q_1(q_1-2k)p_2^2-\half q_2^2p_1^2-\frac{akq_1}{\sqrt{q_1^2+q_2^2}}
-q_2^2k^2c_1 \\
&&+\frac{1}{2k^2}\left(\frac{q_1^2+q_2^2-2q_1k}{q_2}p_2
-\frac{(q_1-k)^2+q_2^2}{(q_1-k)}p_1\right)c_2
+\half\left(\frac{q_2^2p_1}{q_1-k}- q_2p_2\right)c_3
\end{eqnarray*}
and the three Poisson structures are:
\BD
\Pp=\left(
\begin{array}{ccccccc}
0&0&1&0&0&0&0 \\
0&0&0&1&0&0&0\\
-1&0&0&0&0&0&0\\
0&-1&0&0&0&0&0\\
0&0&0&0&0&0&0\\
0&0&0&0&0&0&0\\
0&0&0&0&0&0&0
\end{array}
\right)
\ED
\BD
\Q = \left(
\small{\begin{array}{ccccccc}
0&0&q_1\left(q_1-2k\right)&\left(q_1-k\right)q_2&X_1&0&0\\
0&0&\left(q_1-k\right)q_2&q_2^2&X_2&0&0\\
q_1\left(2k-q_1\right)&\left(k-q_1\right)q_2&0&\left(q_1-k\right)p_2-q_2p_1&
X_3&0&0\\
\left(k-q_1\right)q_2&-q_2^2&\left(k-q_1\right)p_2+q_2p_1&0&X_4&0&0\\
-X_1&-X_2&-X_3&-X_4&0&0&0\\
0&0&0&0&0&0&0\\
0&0&0&0&0&0&0
\end{array}}
\right)
\ED
\BD
\R = \left(
\scriptsize{\begin{array}{ccccccc}
0&0&
\begin{array}{c}
\frac{\left(q_2^2+k^2\right)p_1}{2k^2\left(q_1-k\right)}\\[4pt]
-\frac{q_2p_2}{2k^2}
\end{array}
&\!\!\!\!\!\! \frac{\left(q_1-k\right)p_2-q_2p_1}{2k^2}
&\!\!\!0&Y_1&X_1\\
0&0&\!\!\!\!\!\! \frac{\left(q_1-k\right)p_2-q_2p_1}{2k^2}&
\!\!\!\!\!\! \begin{array}{c}
\frac{\left(q_1-k\right)p_1}{2k^2}-\\[4pt]
\frac{q_1\left(q_1-2k\right)p_2}{2q_2k^2}
\end{array}
&\!\!\!0&Y_2&X_2\\
\begin{array}{c}
\frac{\left(q_2^2+k^2\right)p_1}{2k^2\left(k-q_1\right)}\\[4pt]
+\frac{q_2p_2}{2k^2}
\end{array}
&\frac{\left(k-q_1\right)p_2+q_2p_1}{2k^2}&0&
\!\!\!\!\!\! \begin{array}{c}
\frac{\left(q_1-k\right)^2p_2^2+p_1^2q_2^2}{2q_2k^2\left(q_1-k\right)}\\
-\frac{p_2p_1}{k^2}
\end{array}
&\!\!\!0&Y_3&X_3\\
\frac{\left(k-q_1\right)p_2+q_2p_1}{2k^2}&
\begin{array}{c}
\frac{\left(k-q_1\right)p_1}{2k^2}+\\[4pt]
\frac{q_1\left(q_1-2k\right)p_2}{2q_2k^2}
\end{array}
&
\!\!\!\!\!\! \begin{array}{c}
\frac{\left(q_1-k\right)^2p_2^2
+p_1^2q_2^2}{2k^2q_2\left(q_1-k\right)}\\[4pt]
+\frac{p_2p_1}{k^2}
\end{array}
&0&\!\!\!0&Y_4&X_4\\
0&0&0&0&\!\!\!0&0&0\\
-Y_1&-Y_23&-Y_3&-Y_4&\!\!\!0&0&0\\
-X_1&-X_2&-X_3&-X_4&\!\!\!0&0&0
\end{array}}
\right)
\ED
where $X=X_{\widetilde{H}}=\Pp\D\widetilde{H}$ and $Y=X_{\widetilde{K}}=\Pp\D\widetilde{K}$.

Finally in the case of polar coordinates the transformation between the natural coordinates and the coordinates $\{\lambda_i,\mu_i\}$ is given by:
\BD
\begin{array}{ll}
q_1 = \sqrt{\frac{\lambda_1}{1+\lambda_2^2}}, &
q_2 = \lambda_2\sqrt{\frac{\lambda_1}{1+\lambda_2^2}} \\
p_1 = \frac{2\mu_1\lambda_1-\mu_2\lambda_2^3-\lambda_2\mu_2}{\sqrt{\lambda_1(1+\lambda_2^2)}} &
p_2 = \frac{\mu_2\lambda_2^2+2\lambda_1\lambda_2\mu_1+\mu_2}{\sqrt{\lambda_1(1+\lambda_2^2)}}
\end{array}
\ED
and with the choice
\begin{eqnarray*}
\phi_1(z) &=& -\frac{a}{z}, \\
\phi_2(z) &=& 0
\end{eqnarray*}
the deformed Hamiltonians become
\begin{eqnarray*}
\widetilde{H} &=&
\half (p_1^2+ p_2^2)-\frac{a}{\sqrt{q_1^2+q_2^2}}-(q_1^2+q_2^2)c_1\\
&&-\frac{q_1p_1+q_2p_2-2q_1^2(q_1p_2-q_2p_1)}{2(q_1^2+q_2^2)^2}c_2
-\frac{q_1p_1+q_2p_2}{2(q_1^2+q_2^2)^2}c_3
+\frac{q_2c_4}{q_1(q_1^2+q_2^2)} \\
\widetilde{K} &=&
q_1q_2p_1p_2-\half(q_2^2p_1^2+q_1^2p_2^2)
-q_1^2\frac{q_1p_2-q_2p_1}{q_1^2+q_2^2}c_2-\frac{q_2}{q_1}c_4\,.
\end{eqnarray*}
They are put in the recursion scheme (\ref{ric_pol}) by the three Poisson structures
\BD
\Pp=\left(
\begin{array}{cccccccc}
0&0&1&0&0&0&0&0 \\
0&0&0&1&0&0&0&0\\
-1&0&0&0&0&0&0&0\\
0&-1&0&0&0&0&0&0\\
0&0&0&0&0&0&0&0\\
0&0&0&0&0&0&0&0\\
0&0&0&0&0&0&0&0
\end{array}
\right)
\ED
\BD
\Q = \left(
\scriptsize{\begin{array}{cccccccc}
0&0&q_1^2+\frac{q_2^3}{q_1\left(q_1^2+q_2^2\right)}&
q_1q_2-\frac{q_2^2}{q_1^2+q_2^2}&\!\!\!\!\!\! X_1&\!\!\!\!\!\! 0&\!\!\!\!\!\! 0&\!\!\!\!\!\! Y_1\\
0&0&q_1q_2-\frac{q_2^2}{q_1^2+q_2^2}&
q_2^2+\frac{q_1q_2}{q_1^2+q_2^2}&
\!\!\!\!\!\! X_2&\!\!\!\!\!\! 0&\!\!\!\!\!\! 0&\!\!\!\!\!\! Y_2\\
-q_1^2-\frac{q_2^3}{q_1\left(q_1^2+q_2^2\right)}&
\frac{q_2^2}{q_1^2+q_2^2}-q_1q_2&0&
\!\!\!\!\!\! \begin{array}{c}
q_1p_2-q_2p_1 - \\
\frac{q_2(q_1p_2-q_2p_1)}
{q_1\left(q_1^2+q_2^2\right)}
\end{array}
&\!\!\!\!\!\! X_3&\!\!\!\!\!\! 0&\!\!\!\!\!\! 0&\!\!\!\!\!\! Y_3\\
\frac{q_2^2}{q_1^2+q_2^2}-q_1q_2&
-q_2^2-\frac{q_1q_2}{q_1^2+q_2^2}&
\!\!\!\!\!\! \begin{array}{c}
q_2p_1-q_1p_2 - \\
\frac{q_2(q_2p_1-q_1p_2)}
{q_1\left(q_1^2+q_2^2\right)}
\end{array}
&0&\!\!\!\!\!\! X_4&\!\!\!\!\!\! 0&\!\!\!\!\!\! 0&\!\!\!\!\!\! Y_4\\
-X_1&-X_2&-X_3&-X_4&\!\!\!\!\!\! 0&\!\!\!\!\!\! 0&\!\!\!\!\!\! 0&\!\!\!\!\!\! 0\\
0&0&0&0&\!\!\!\!\!\! 0&\!\!\!\!\!\! 0&\!\!\!\!\!\! 0&\!\!\!\!\!\! 0\\
0&0&0&0&\!\!\!\!\!\! 0&\!\!\!\!\!\! 0&\!\!\!\!\!\! 0&\!\!\!\!\!\! 0\\
-Y_1&-Y_2&-Y_3&-Y_4&\!\!\!\!\!\! 0&\!\!\!\!\!\! 0&\!\!\!\!\!\! 0&\!\!\!\!\!\! 0
\end{array}}
\right)
\ED
\BD
\R = \left(
\small{\begin{array}{cccccccc}
0&0&A&B&0&Y_1&X_1&0\\
0&0&B&C&0&Y_2&X_2&0\\
-A&-B&0&D&0&Y_3&X_3&0\\
-B&-C&-D&0&0&Y_4&X_4&0\\
0&0&0&0&0&0&0&0\\
-Y_1&-Y_2&-Y_3&-Y_4&0&0&0&0\\
-X_1&-X_2&-X_3&-X_4&0&0&0&0\\
0&0&0&0&0&0&0&0
\end{array}}
\right)
\ED
where $X=X_{\widetilde{H}}=\Pp\D\widetilde{H}$, $Y=X_{\widetilde{K}}=\Pp\D\widetilde{K}$ and
\begin{eqnarray*}
A &=& q_1^2\,\frac{q_1p_1+q_2p_2+2q_2^2(q_1p_2-q_2p_1)}{2\left(q_1^2+q_2^2\right)^2} \\
B &=& q_1q_2\,\frac{q_1p_1+q_2p_2-2q_1^2(q_1p_2-q_2p_1)}{2\left(q_1^2+q_2^2\right)^2} \\
C &=& \frac{q_2^2(q_1p_1+q_2p_2)+2q_1^4(q_1p_2-q_2p_1)}{2\left(q_1^2+q_2^2\right)^2} \\
D &=& (q_1p_2-q_2p_1)\frac{q_1p_1+q_2p_2-2q_1^2(q_1p_2-q_2p_1)}{2\left(q_1^2+q_2^2\right)^2}
\end{eqnarray*}

\section{Final remarks}
In this article a procedure to construct trihamiltonian extensions of classical separable systems has been presented. This procedure has been tested in the particularly simple case of the Euclidean plane, but it is in principle applicable to more general cases. It is articulated in the following steps:
\begin{enumerate}
\item Write down the separation relations of a given system with $n$ degree of freedom in the form involving the St\"ackel matrix of  the separation coordinates:
\BD
\left(
\begin{array}{ccc}
a_{11}(q_1) & \cdots & a_{1n}(q_1) \\
a_{21}(q_2) & \cdots & a_{2n}(q_2) \\
\vdots && \vdots \\
a_{n1}(q_n) & \cdots & a_{nn}(q_n)
\end{array}
\right)
\left(
\begin{array}{c}
K_1(\{p_i,q_i\}) \\
K_2(\{p_i,q_i\}) \\
\vdots \\
K_n(\{p_i,q_i\})
\end{array}
\right)=\left(
\begin{array}{c}
\Phi_1(p_1,q_1) \\
\Phi_2(p_2,q_2) \\
\vdots \\
\Phi_n(p_n,q_n)
\end{array}
\right)
\ED
\item Transform, through a transformation to a suitable system of coordinates $\{\lambda_i,\mu_i\}$, and in some case a linear combination of the Hamiltonians,  the St\"ackel relation in such a way that the $i$-th row contains just ordered powers of $\lambda_i$. This is the most troublesome step, in fact it's not clear which kind of St\"ackel matrix can be  put in this form, although it is possible for all the eleven orthogonal separable webs in $\Reali^3$.
\item Now the separation relation encoded by the $i$-th row of St\"ackel matrix is a polynomial in the coordinate $\lambda_i$, its coefficients are a set of Hamiltonians $H_j$ obtained from the recombination of the $K_j$. These polynomial can be grouped on the basis of their form: to each different form of the polynomials it is associated a trihamiltonian recursion scheme between the $H_j$ and a corresponding polynomial function. In general there will be $m\le n$ of such functions, labelled $f_1,\ldots,f_m$. The coefficients of the polynomial functions $f_l$ are the deformed Hamiltonians $\widetilde{H}_j$, together with a suitable number of Casimir functions.
\item Solving the $n$ equations (linear in the $n$ deformed Hamiltonians)
\BD
\left\{
\begin{array}{rcl}
f_1(\lambda_1;\{\widetilde{H}_j\}) &=& r_1(\lambda_1,\mu_1) \\
&\vdots\\
f_m(\lambda_n;\{\widetilde{H}_j\}) &=& r_n(\lambda_n,\mu_n)
\end{array}
\right.
\ED
the explicit forms of the deformed Hamiltonians is obtained.
\item Extending in a trivial way the canonical Poisson tensor, the Poisson tensor $\Pp$ on the extended space is obtained. Instead, from the deformed Hamiltonian and the recursion scheme, it is possible to construct the two vector fields $\Xq$ and $\Xr$, and then the two tensors $\Q=\Qd-\DLie{\Xq,\Pp}$ and $\R=\Rd-\DLie{\Xr,\Pp}$.
\item The \emph{conjecture}, on which this procedure is based, is that the three tensors $\Pp$, $\Q$ and $\R$ constructed in this way are all Poisson tensors mutually compatible. Moreover the $m$ functions $f_l(\lambda,\mu)$ are common Casimir functions for the two Poisson pencils $\Q-\lambda\Pp$ and $\R-\mu\Pp$. As a consequence the deformed Hamiltonians $\widetilde{H}_j$ satisfy the required trihamiltonian recursion scheme.
\end{enumerate}

\section*{Acknowledgements}
I wish to thank prof. Guido Magnano and dr. Claudia Chanu for the pieces of advice and the suggestions they gave me during the preparation of this paper.

\end{document}